\documentclass[a4paper]{jpconf}
\usepackage{graphicx}
\begin{document}
\title{Measurement of beauty-hadron decay electrons in Pb--Pb collisions at $\sqrt{s_{NN}}=2.76 ~\textrm{TeV}$ with ALICE}

\author{Martin V\"{o}lkl for the ALICE Collaboration}

\address{Physikalisches Institut der Universit\"{a}t Heidelberg, Heidelberg, Germany}

\ead{voelkl@physi.uni-heidelberg.de}
\begin{abstract}
The ALICE Collaboration at the LHC studies heavy-ion collisions to investigate the properties of the Quark-Gluon Plasma (QGP). Heavy quarks (charm and beauty) are effective probes for this purpose. Both their energy loss in the medium as well as their possible thermalization yield information about the medium properties. Experimentally, the reconstruction of hadrons with charm valence quarks is possible. For hadrons with beauty valence quarks a promising strategy is the measurement of their decay electrons. To separate these from the background electrons (mainly from charm hadron decays, photon conversions or light-meson decays) the large decay length of beauty hadrons can be utilized. It leads to a relatively large typical impact parameter of the decay electrons. By comparing the impact parameter distribution of the signal electrons with those from the background sources, the signal can be statistically separated from the background. For this purpose a maximum likelihood fit is employed using impact parameter distribution templates from simulations. The resulting nuclear modification factor for electrons from beauty-hadron decays shows a sizeable suppression for $p_\mathrm{T}>3~ \mathrm{GeV}/c$, albeit still with large uncertainties.
\end{abstract}
\vspace{-0.5cm}
\section{Introduction}
The measurement of hadrons with heavy valence quarks (charm or beauty) via their decay products in heavy-ion collisions can give insight into the properties of the Quark-Gluon Plasma. Of particular interest are the energy loss the quarks experience in the medium and their possible thermalization. Theoretical calculations predict a lower energy loss for quarks than for gluons and a lower energy loss for heavy quarks compared with lighter ones making a flavor separated measurement interesting \cite{Dokshitzer2001199}. Such a measurement is possible through the observation of hadrons containing heavy valence quarks.

\begin{figure}[]
\begin{minipage}{0.45\textwidth}
\includegraphics[width=0.90\textwidth]{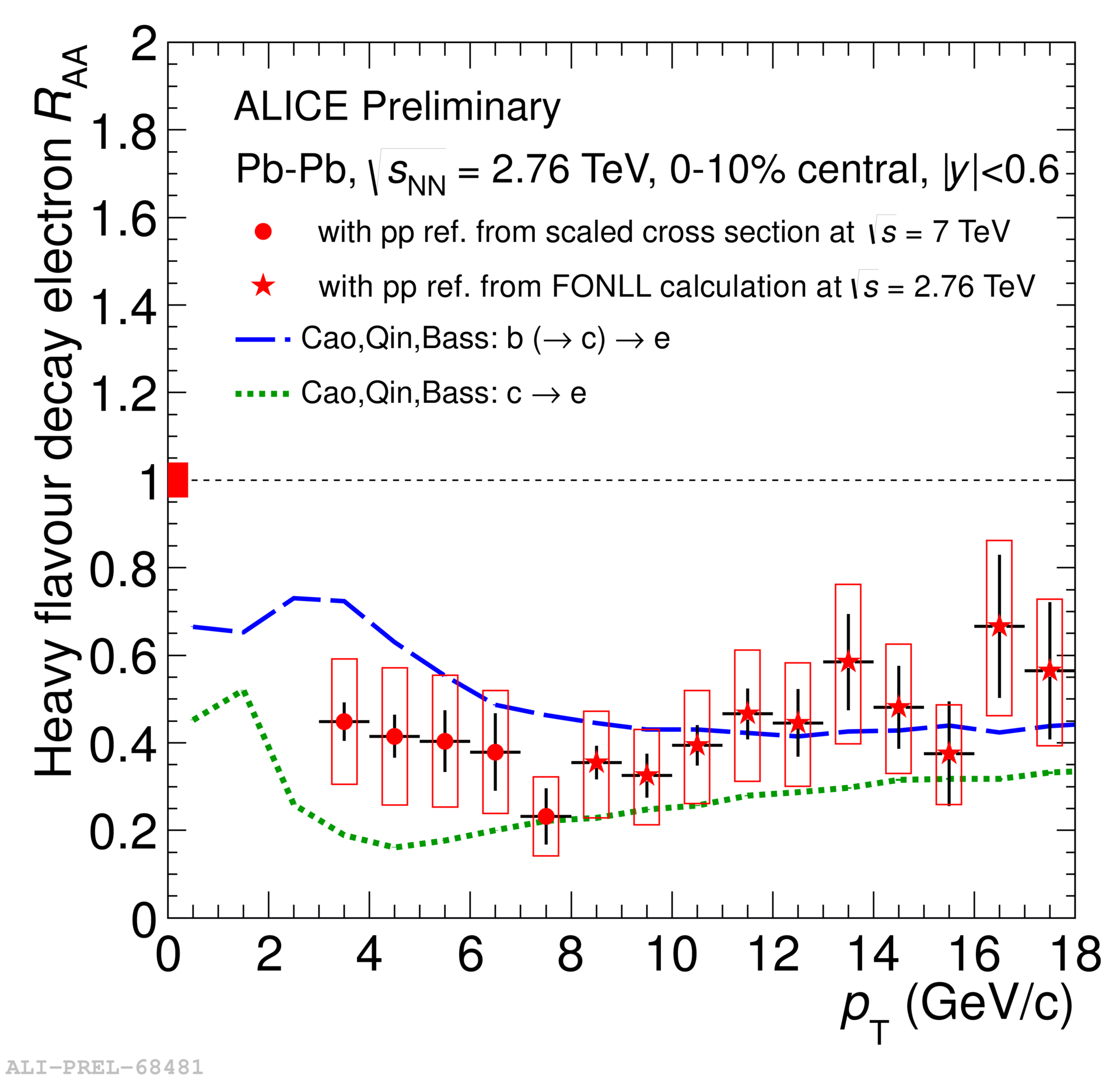}
\caption{\label{HFEEMCal}$R_{\mathrm{AA}}$ of electrons from heavy-flavor decays in central Pb--Pb collisions \cite{Sakai2013661c}.}
\end{minipage}\hspace{0.05\textwidth}%
\begin{minipage}{0.52\textwidth}
\includegraphics[width=0.99\textwidth]{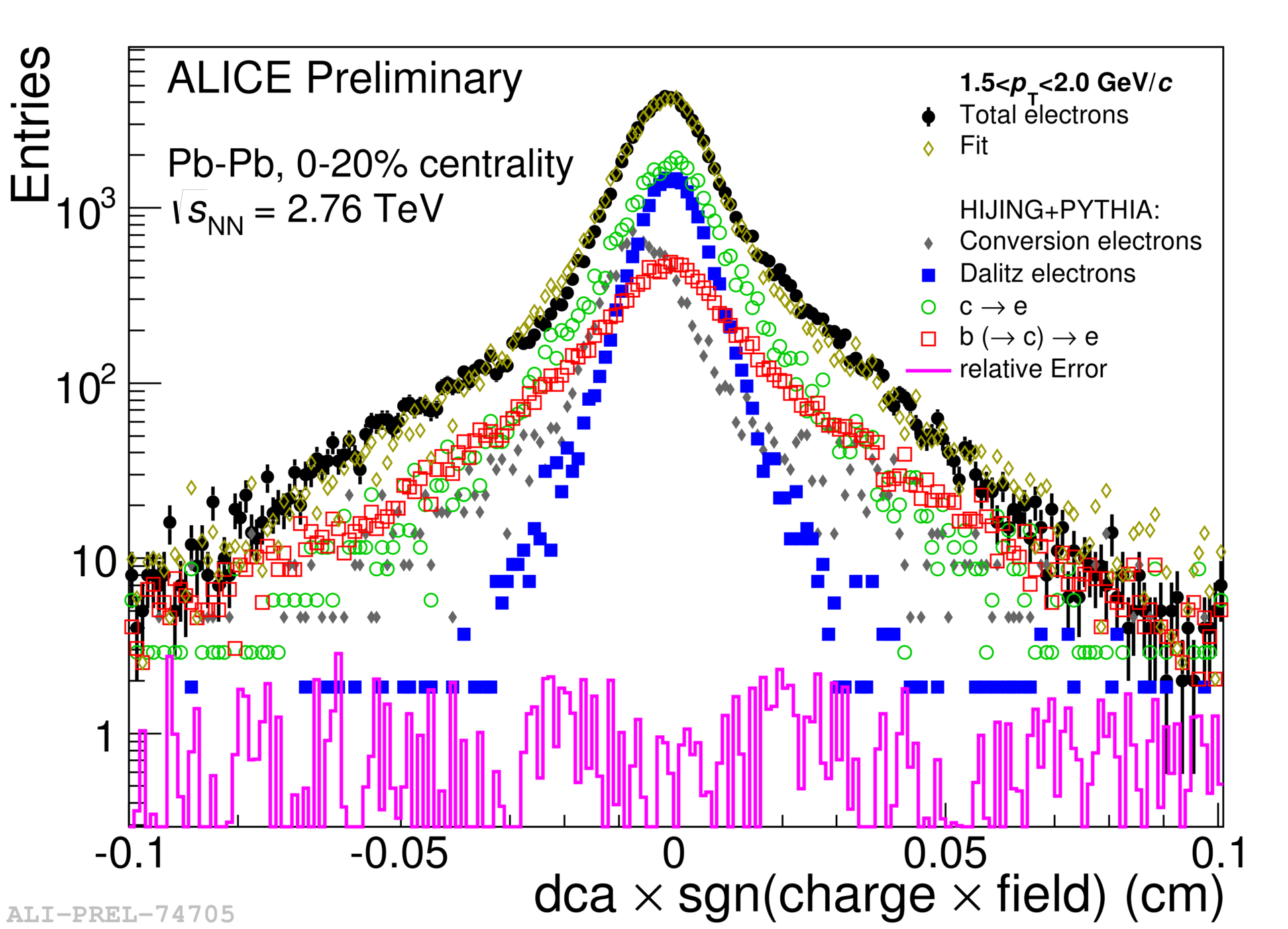}
\caption{\label{DCADist}Impact parameter distribution of electron candidates in data and electrons from different sources in simulations.}
\end{minipage}
\end{figure}

For charm hadrons a full reconstruction from hadronic decay products is feasible \cite{raey}. Due to small branching ratios this approach is not used for the decays of beauty hadrons. To access beauty quarks, use can be made of the large branching ratios ($\approx 10\%$) of decays with an electron (or positron) in the final state. A comparison of the $p_{\mathrm{T}}$ distributions of electrons from beauty-hadron decays in Pb--Pb and pp collisions based on the nuclear modification factor ($R_{\mathrm{AA}}$) is sensitive to the medium effects on b-quarks. $R_{\mathrm{AA}}$ is the ratio of the $p_{\mathrm{T}}$-differential yield in Pb--Pb collisions to the corresponding cross section in pp collisions scaled with the nuclear overlap function. \Fref{HFEEMCal} shows a measurement of the nuclear modification factor of electrons from heavy-flavor hadron decays for the $10\%$ most central Pb--Pb collisions at $\sqrt{s_{NN}}=2.76 ~\textrm{TeV}$ \cite{Sakai2013661c}. The value is significantly below unity, pointing to an energy loss of heavy quarks in the hot and dense medium. As shown by the model expectations \cite{CaoPaper,CaoPrivateComm}, it is interesting to know what the contributions from beauty and charm are separately.

\section{Electron Identification and Background Sources}
Electron identification on the data set of Pb--Pb collisions recorded in 2010 is done using the ALICE Time Projection Chamber (TPC) and the Time of Flight detector (TOF) \cite{PERFORMANCE}. Particle identification with the TPC is based on the specific energy loss of charged particles in its large gas volume. Electrons are identified by requiring the measured time-of-flight to be within $3 \sigma$ of the expected value for electrons and the $\mathrm{d}E/\mathrm{d}x$ in the TPC to be within $0$ and $+3 \sigma$ with respect to the expected $\mathrm{d}E/\mathrm{d}x$ of electrons. This asymmetric cut keeps the contamination due to pions low. The remaining hadron contamination was studied and was found not to contribute strongly to the systematic uncertainty of the measured spectrum of beauty-hadron decay electrons.

The number of electrons from sources other than heavy-flavor decays can be estimated using other measurements performed by the ALICE collaboration. The background is dominated by Dalitz decays of light mesons and photon conversions in the detector material.

Knowledge of this background allows for a measurement of the number of electrons from heavy-flavor decays based on the subtraction of background electrons from the total number of electrons \cite{PhysRevD.86.112007}. The measured D-meson $p_{\mathrm{T}}$-differential cross sections allow for the subtraction of the charm-decay contribution by simulating the decay of the charm hadrons to electrons. However, a measurement of the beauty contribution with this approach leads to large uncertainties because a large background is subtracted from a large electron sample. Thus, additional information is needed to improve the precision of the measurement.

\section{The Impact Parameter}
The typically large decay length of beauty hadrons \cite{PDG} can be used for the separation of the different electron sources. Due to the fact that only one decay particle is measured (the electron), the decay vertex cannot be measured directly. The proxy quantity is the track impact parameter. A track propagated backwards from a secondary vertex will typically not go through the primary vertex. The distance of closest approach to the primary vertex of this backwards propagated track is called impact parameter. Its sign can be positive or negative depending on the position of the primary vertex relative to the propagated track. An accurate measurement of the impact parameter is possible due to the excellent resolution of the tracks and the primary vertex provided by the Inner Tracking System (ITS) \cite{PERFORMANCE}. Four types of electron sources can be distinguished via their impact parameter distributions as shown in \Fref{DCADist}:
\begin{description}
\item[Electrons from beauty-hadron decays.] Due to the large decay length ($c \tau \approx 500 ~\mathrm{\mu m}$) of beauty hadrons this distribution is the widest among the different electron sources.
\item[Electrons from charm-hadron decays.] The smaller decay length of the charm hadrons ($c \tau \approx 100-300~ \mathrm{\mu m}$) leads to a narrower distribution as compared to beauty-hadron decay electrons.
\item[Electrons from the primary vertex.] All electrons produced at or very close to the primary vertex have the same impact parameter distribution which is determined by the resolution of the track parameters. These electrons mostly come from Dalitz decays of $\pi^{0}$ and will be referred to as Dalitz electrons.
\item[Electrons from photon conversions in the detector material.] Due to the small opening angle of photon conversions a non-zero impact parameter is due to the influence of the magnetic field. It has proven useful to multiply the impact parameter by the signs of the particle charge and field polarity. This results in an asymmetric distribution.
\end{description}
There are two approaches of using the impact parameter information to measure electrons from beauty-hadron decays. The first approach is based on a cut on the minimum absolute value of the impact parameter in order to preferentially select beauty-hadron decay electrons. Afterwards the remaining background electrons are subtracted. The impact parameter cut allows for a substantial improvement of the signal-to-background ratio leading to a smaller uncertainty on the beauty-hadron decay electron yield with respect to the case in which the impact parameter selection is not applied. The pp reference at $\sqrt{s}=7~\mathrm{TeV}$ for the $R_{\mathrm{AA}}$ \cite{Abelev201313} was measured with this approach. The second approach is based on the statistical separation of the electron sources. It was used for the Pb--Pb analysis as described in the next section.

\section{The Impact Parameter Fit Method}
The basic idea is to compare the impact parameter distributions of the various electron sources from simulation (templates) to the impact parameter distribution of all measured electrons to estimate the individual contributions. The simulations use the particles from event generators (HIJING \cite{HIJING}, PYTHIA \cite{PYTHIA}) which are transported through the ALICE apparatus with GEANT3 \cite{GEANT}. The accuracy of the simulated impact parameter distributions was shown to not be a dominant source of systematic uncertainty. The fit of the impact parameter distributions is performed separately for each of the six $p_{\mathrm{T}}$-intervals considered in this analysis. With the template distributions $a_{source,bin}$, the expectation values of the counts in each impact-parameter bin are introduced as the free fit parameters $A_{source,bin}$. The expectation value of the data counts is the fit function $f_{bin}$. It can be written as \cite{Barlow1993219}:
\[f_{bin} = \sum_{source} p_{source} \cdot A_{source,bin} ~.\]
The $p_{source}$ are the strength factors of the fit.
Assuming Poissonian fluctuations in the data bins $d_{bin}$ and template bins $a_{source,bin}$, this leads to the likelihood:
\[\log L = \sum_{bin} d_{bin} \log f_{bin} - f_{bin} + \sum_{bin}\sum_{source} a_{source,bin} \log A_{source,bin} - A_{source,bin} ~,\]
neglecting constant terms. The first term incorporates the fluctuations of the data, as in a fit with a smooth function. The second term includes the fluctuations of the template bin contents.
The minimization of the $A_{source,bin}$ can be done fairly efficiently \cite{Barlow1993219}, leaving only the amplitudes $p_{source}$ as free parameters for numerical minimization. An example of such a fit can be found in \Fref{DCADist}.

To estimate the uncertainties from the maximum likelihood method, toy data and toy simulation distributions were created from the templates. Taking the measured results as a starting point the fit was repeatedly simulated. The spread of the resulting amplitudes was identified as the statistical uncertainty while the difference of the mean of the results to the true value of the toy model was interpreted as a systematic bias of the maximum likelihood method.

\begin{figure}[]
\begin{minipage}{0.47\textwidth}
\includegraphics[width=0.99\textwidth]{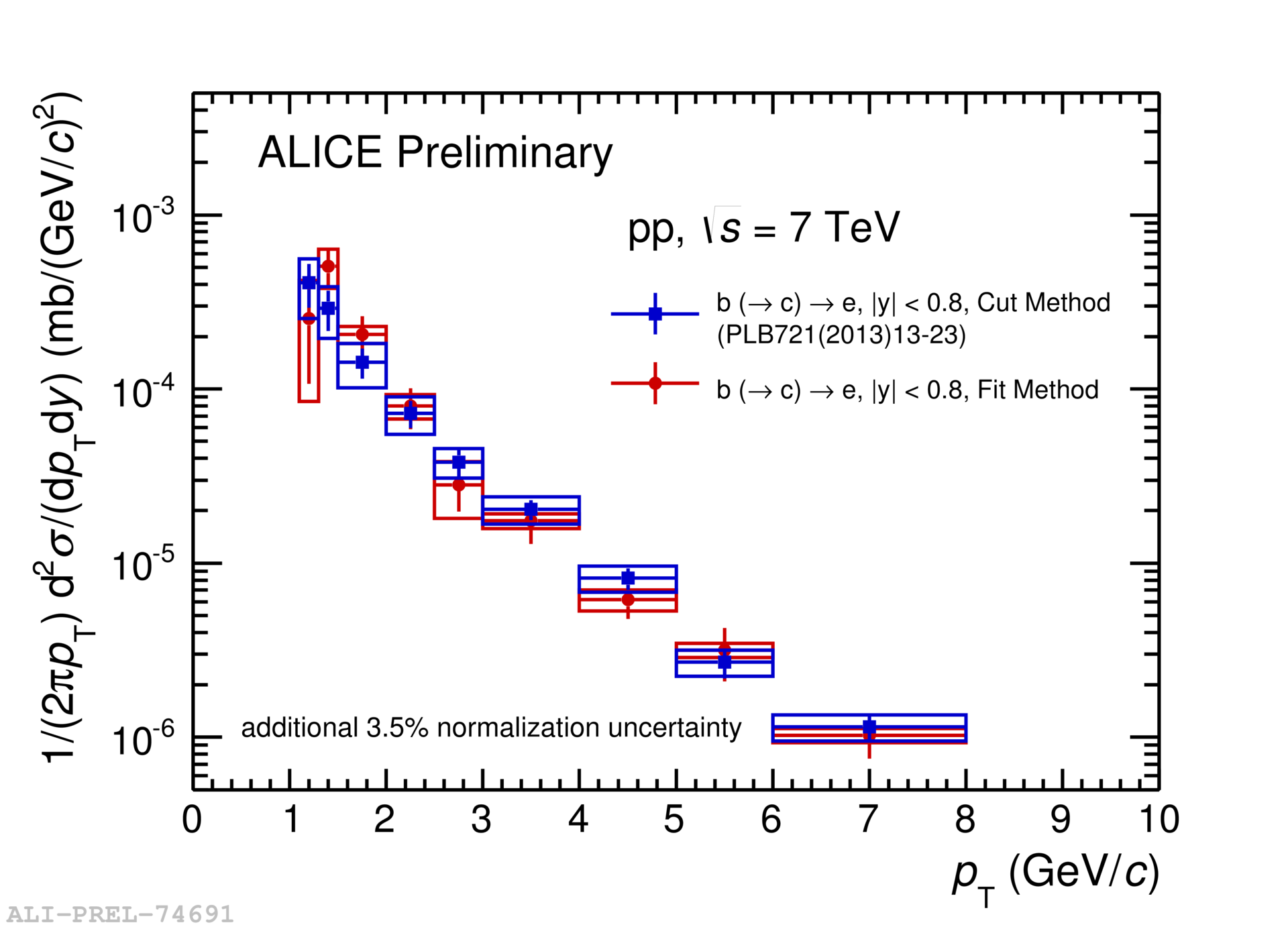}
\caption{\label{ppSpec}Comparison of beauty-decay electron cross sections obtained from fit and cut methods in pp collisions at $\sqrt{s}=7~\mathrm{TeV}$.}
\end{minipage}
\hspace{0.05\textwidth}%
\begin{minipage}{0.47\textwidth}
\includegraphics[width=0.99\textwidth]{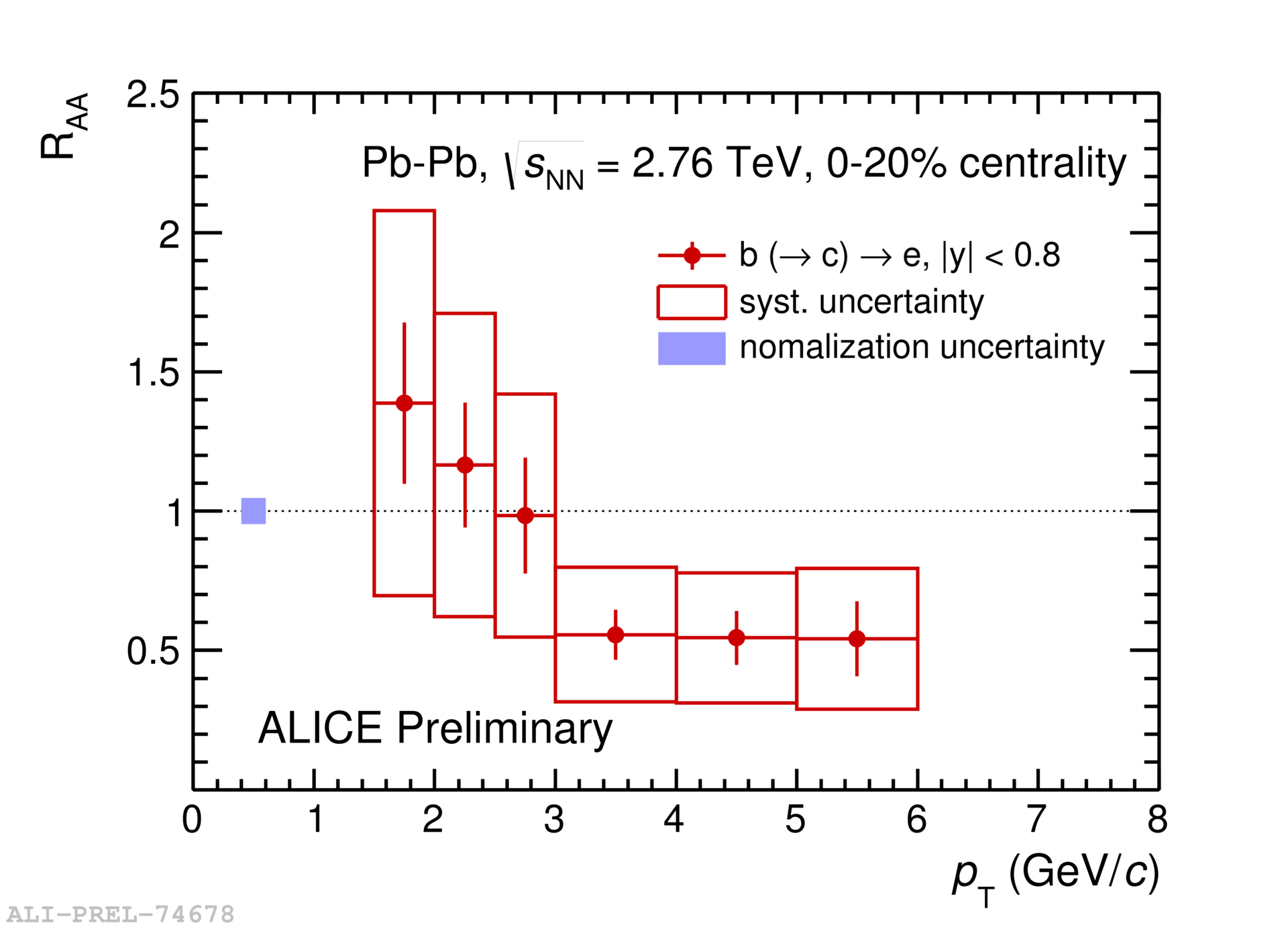}
\caption{\label{RAA}$R_{\mathrm{AA}}$ of electrons from beauty-hadron decays measured in central Pb--Pb collisions as a function of $p_{\mathrm{T}}$.}
\end{minipage} 
\end{figure}

\section{Results}
The method was tested on the data sample of pp collisions at $\sqrt{s}=7 ~\textrm{TeV}$ and compared with the results from the impact parameter cut method. The results from the two methods are in good agreement within their systematic uncertainties as shown in \Fref{ppSpec}. The systematic uncertainties of the two approaches are of a similar size. The advantage of the fit method is its independence from other measurements. The fit method was applied to the 20\% most central Pb--Pb collisions at $\sqrt{s_{NN}}=2.76 ~\textrm{TeV}$ recorded in 2010. The uncertainties are currently dominated by the strong selection criteria of the particle identification in Pb--Pb collisions with smaller contributions from the accuracy of the simulated impact parameter distributions, the pp reference and its energy scaling and the fit method. The resulting $R_{\mathrm{AA}}$ is shown in \Fref{RAA}. It shows a suppression ($R_{\mathrm{AA}}<1$) for $p_\mathrm{T}>3~ \mathrm{GeV}/c$ - albeit with sizeable uncertainties - with the indication of a rise towards lower $p_{\mathrm{T}}$. This is consistent with in-medium energy loss of beauty quarks causing a redistribution of the beauty-hadron decay electrons towards lower $p_{\mathrm{T}}$ in Pb--Pb collisions compared to pp collisions. Due to the large uncertainties no clear indication of a quark-mass dependent in-medium energy loss is observed at the current stage of the analysis.

\section*{References}
\bibliography{procBib}

\end{document}